\documentclass{article}

\usepackage{arxiv}

\usepackage{amsmath}
\usepackage[colorlinks=true, allcolors=blue]{hyperref}
\usepackage{color}

\usepackage[utf8]{inputenc} % allow utf-8 input
\usepackage[T1]{fontenc}    % use 8-bit T1 fonts
\usepackage{hyperref}       % hyperlinks
\usepackage{url}            % simple URL typesetting
\usepackage{booktabs}       % professional-quality tables
\usepackage{amsfonts}       % blackboard math symbols
\usepackage{nicefrac}       % compact symbols for 1/2, etc.
\usepackage{microtype}      % microtypography
\usepackage{lipsum}
\usepackage{graphicx}
\graphicspath{ {./images/} }

\title{A Machine Learning Approach to Detecting Albedo Anomalies on the Lunar Surface Using Data-Driven Methods}

\author{
 Sofia Strukova \\
  Department of Information and Communication Engineering\\
  University of Murcia\\
  Murcia, Spain 30100 \\
  \texttt{strukovas@um.es} \\
   \And
 Sergei Gleyzer \\
  Department of Physics \& Astronomy\\
  University of Alabama\\
  Tuscaloosa, AL 35401 \\
  \texttt{sgleyzer@ua.edu} \\
  \And
 Patrick Peplowski \\
  Johns Hopkins University Applied Physics Laboratory\\
  Laurel, MD 20723 \\
  \texttt{Patrick.Peplowski@jhuapl.edu} \\
  \And
 Jason P. Terry \\
  Department of Physics \& Astronomy\\
  University of Georgia\\
  Athens, GA 30602 \\
  \texttt{jpterry@uga.edu} \\
}

\begin{document}
\maketitle
\begin{abstract}
This study introduces a data-driven approach using machine learning (ML) techniques to explore and predict albedo anomalies on the Moon's surface. The research leverages diverse planetary datasets, including high-spatial-resolution albedo maps and element maps (LPFe, LPK, LPTh, LPTi) derived from laser and gamma-ray measurements. The primary objective is to identify relationships between chemical elements and albedo, thereby expanding our understanding of planetary surfaces and offering predictive capabilities for areas with incomplete datasets. To bridge the gap in resolution between the albedo and element maps, we employ Gaussian blurring techniques, including an innovative adaptive Gaussian blur. Our methodology culminates in the deployment of an Extreme Gradient Boosting Regression Model, optimized to predict full albedo based on elemental composition. Furthermore, we present an interactive analytical tool to visualize prediction errors, delineating their spatial and chemical characteristics. The findings not only pave the way for a more comprehensive understanding of the Moon's surface but also provide a framework for similar studies on other celestial bodies.
\end{abstract}

% keywords can be removed

\section{Introduction}

The study of planetary surfaces provides a window into the geologic history of celestial bodies, allowing scientists to piece together the evolutionary stories of planets and moons. The tools used to study these surfaces vary from telescopes on Earth to intricate instruments aboard spacecrafts. Through electromagnetic wavelengths, ranging from radar to gamma-ray, the mysteries of the surfaces are unveiled, each wavelength offering a different perspective. Notably, these wavelengths shed light on the surface's chemistry, mineralogy, and past events. The Moon, for instance, has been extensively observed and its albedo has been correlated with the presence of specific elements like iron. However, when it comes to other celestial bodies, the relationships between observations may not be as well-established, and the datasets can be incomplete. Thus, a challenge of accurately predicting the properties of a planetary surface with limited information arises.

Planetary surface observations can be studied through a wide range of electromagnetic wavelengths (e.g., radar, infrared, optical, ultraviolet, x-ray, gamma-ray). Each wavelength provides unique information about the surface's chemistry, mineralogy, and history. This includes reflected solar light, the magnitude of which is known as surface albedo. Yet, the information is not entirely independent. For example, the chemical element iron, which is mapped with x-rays and gamma rays, is highly related to optical albedo on the Moon. Knowing this, we can develop high-spatial-resolution predictive maps of iron based on optical data. On other planets, the relationships between the observations are less well-known, and some datasets are missing. We seek to study planetary surfaces by inputting maps of surfaces at all wavelengths available to discover the relationships between the measurements and to make predictions about chemistry that are not directly sampled by observations. This provides a way of studying the geologic history of a planet with existing data, which is valuable given the infrequent opportunities for new measurements by planetary spacecraft.

Many different elements contribute to varying extent to the albedo and the albedo can be related to the underlying elemental distribution. These relationships are generally well-known for the Moon, making it an ideal location to test new approaches to linking measurements at different wavelengths in anticipation of applying these techniques to other planetary surfaces where the relationship is less understood. Thus, we built a predictive model based on the compositional maps of the Moon derived from Lunar Prospector data to quantify the relationships between chemical element concentrations and the surface albedo. Our goals are to make predictions about albedo based on the elemental composition and uncover any possible anomalies that can have roots in the geologic history of a planet.   This provides a way of studying the Moon with existing data, which is valuable given the infrequent opportunities for new measurements by planetary spacecraft. Moreover, the shape and the location of anomalies on the lunar surface can provide insights into processes that produced them and their subsequent evolution.

The remainder of this paper is structured as follows. In Section~\ref{sec:methodology}, we present our research methodology. Our findings are outlined in Section~\ref{sec:results}, while we extend the results, present interpretations, suggestions for future work, and limitations in Section~\ref{sec:discussion}. Finally, we draw our conclusions and future research directions in Section~\ref{sec:conclusions}.

\section{Methodology}
\label{sec:methodology}

\begin{figure*}[!ht]
\includegraphics[width=1\textwidth]{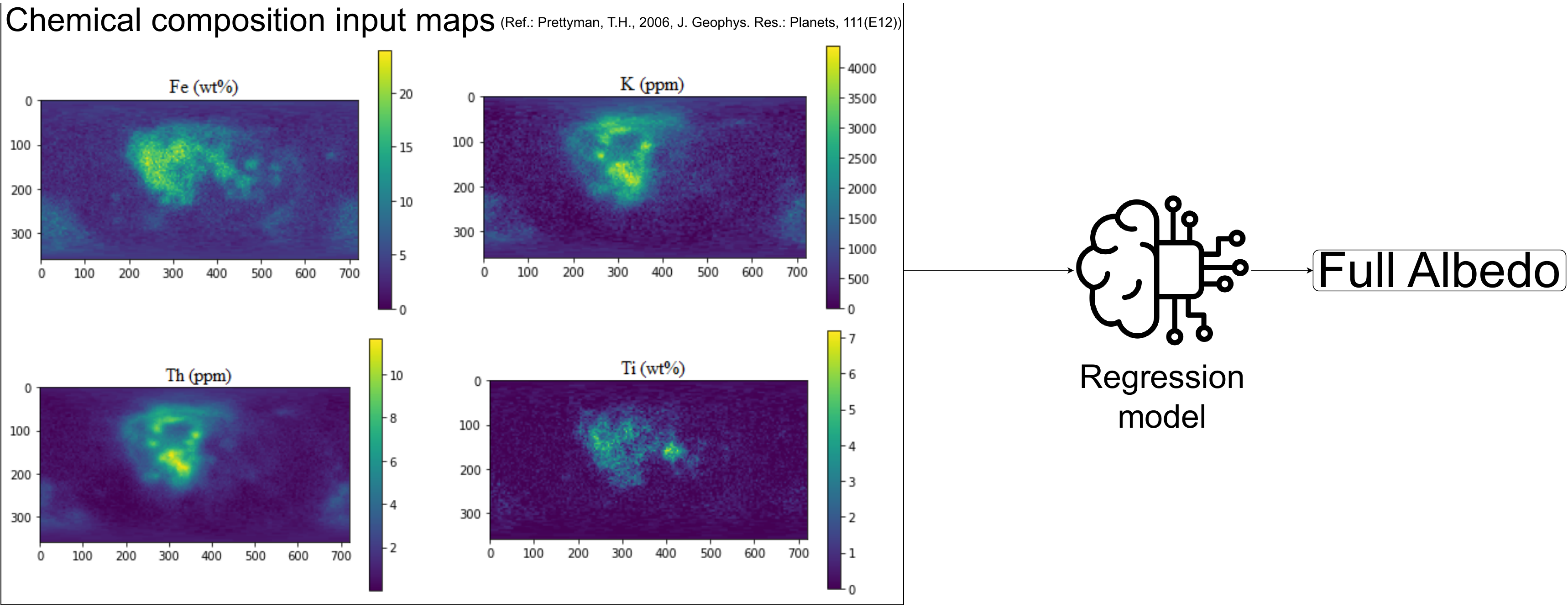}
\centering
\caption{Overview of the methodology to predict the full albedo}
\label{fig:methodology}
\end{figure*}

%FIGURE TO BE UPDATED OR DELETED. 
We will pursue the methodology process presented in Figure~\ref{fig:methodology}. 

%https://drive.google.com/file/d/1DBAdIZsVs-Ilm7enKl_1DGkr6WLaOXSQ/view?usp=sharing

%https://iopscience.iop.org/article/10.3847/1538-4357/aca477/pdf

In our research, we initially focus on the classification of two distinct datasets: albedo maps sourced from high spatial resolution laser measurements and element maps derived from gamma-ray measurements. The dataset consists of several maps of the Moon from the Lunar Prospector Gamma Ray and Neutron Spectrometer~\cite{Prettyman2006}. The element maps include gamma-ray derived concentrations for Fe (iron), K (potassium), Th (thorium), and Ti (titanium). The albedo map was derived from the Lunar Reconnaissance Orbiter LOLA instrument~\cite{Smith2009}. These datasets inherently possess disparate spatial resolutions, with the former measured in the range of tens of meters and the latter possessing an intrinsic spatial resolution of approximately 45 kilometers.

To address this disparity in resolutions, we conduct our experiment under two specific settings: one involving blurring techniques and one without. In the setting that incorporates blurring, we implement two types of Gaussian blur — standard and adaptive. The application-specific determination of the variance/covariance matrix parameters for the Gaussian filters represents a meticulous aspect of this phase.

Next, we turn our attention to the quantitative evaluation of these blurring techniques. For this purpose, we employ metrics such as the R2 score and the root-mean-square error. Based on this comprehensive assessment, we identify optimal parameters for our blurring process, with sigma set to 9 and kSize at (0,0).

Following the resolution harmonization and blurring evaluation, we proceed to train a machine learning (ML) regression model. For this task, we utilize a subset of the lunar surface as our training data, treating chemical elements as input variables and the albedo as the target variable. The Extreme Gradient Boosting Regression Model emerges as the best-performing ML model, configured with specific hyperparameters - namely, a learning objective of regression with squared loss, a learning rate of 0.1, 30 estimators, and a maximum depth of 5.

With this trained model in hand, we generate predictions for the full albedo map, covering both the original and blurred versions. In parallel, we develop an interactive analytical tool designed to enhance our research insights. This tool provides real-time contour mapping functionalities for regions where the predictive error exceeds a predefined threshold. It also allows for multiple user-configurable parameters, including error thresholds and display configurations.

Through this multifaceted research methodology, we aim to establish a robust and nuanced understanding of the variables affecting lunar albedo, thus setting the stage for future research and lunar explorations.

\section{Results}
\label{sec:results}

\subsection{Blurring}

The need for the implementation of image blurring arises from the types of datasets used in this study. Specifically, the disparity in the spatial resolution between the two primary sources of data necessitated an approach to reconcile these differences and facilitate more effective comparative analysis.

Our study relies on two distinct categories of datasets. On the one hand, we utilize albedo maps derived from laser measurements featuring an exceptionally high spatial resolution on the order of tens of meters. Conversely, the elemental maps at our disposal, derived from gamma-ray measurements, offer an intrinsic spatial resolution of approximately 45 kilometers. The drastic incongruity in spatial resolutions between these datasets presents a challenge for direct comparison and analysis.

\begin{figure*}[!ht]
\includegraphics[width=1\textwidth]{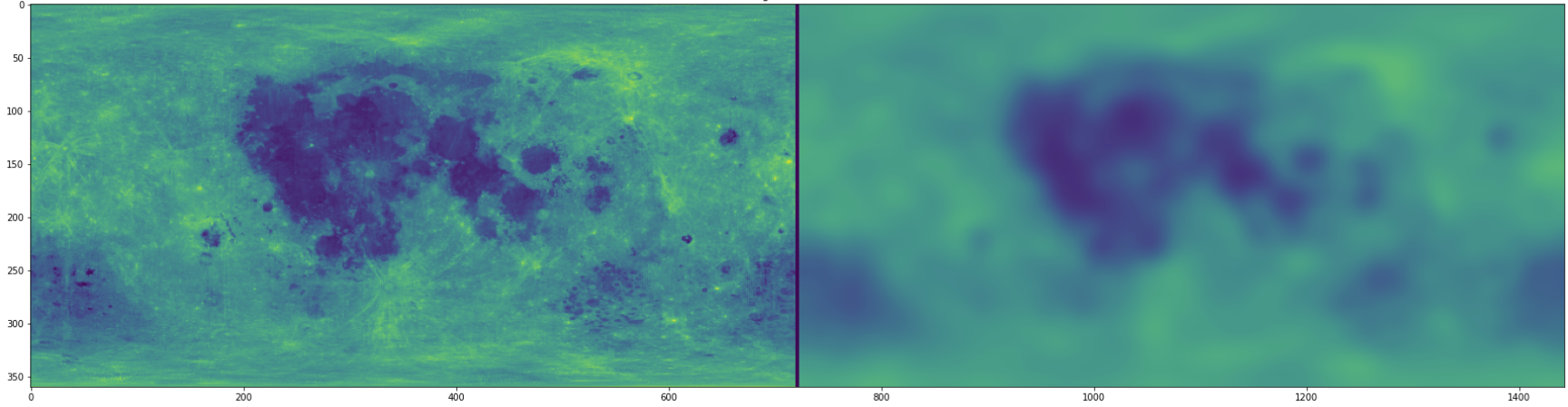}
\centering
\caption{Original (in the left) and blurred (in the right) images of the Moon}
\label{fig:original_blurred}
\end{figure*}

To address this issue, we conducted experiments in two settings: one incorporating image blurring and the other devoid of it (see Figure~\ref{fig:original_blurred}). This decision was geared toward achieving comparability between the disparate datasets. Furthermore, the technique facilitated the development of a ``resolution-tolerant'' ML model that is equipped to minimize potential artifacts stemming from variances in resolution — a significant stride toward the ultimate objective of identifying any concealed resolution artifacts.

\begin{figure*}[!ht]
\includegraphics[width=1\textwidth]{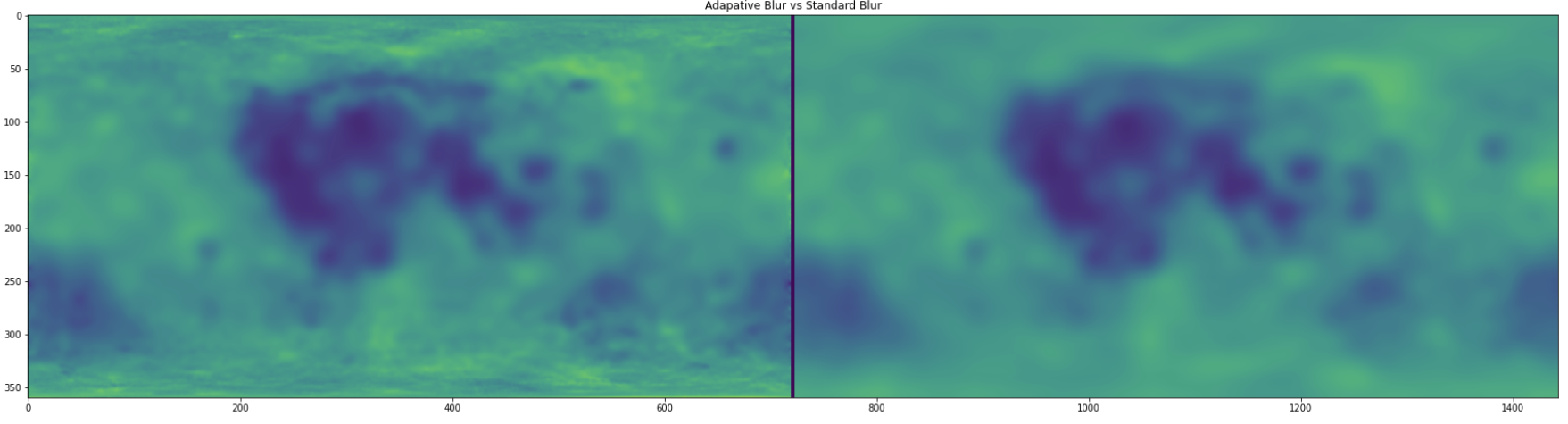}
\centering
\caption{The difference between the adaptive and the standard blurring}
\label{fig:adaptive_blurring}
\end{figure*}

\begin{figure*}[!ht]
\includegraphics[width=1\textwidth]{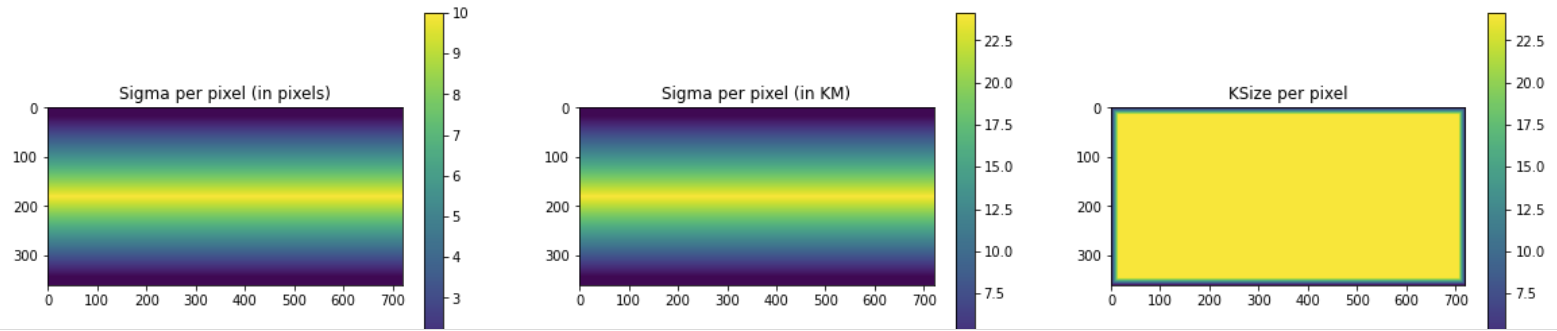}
\centering
\caption{Maps of the amount of sigma applied to each pixel of the image}
\label{fig:sigma}
\end{figure*}

Moreover, the spatial resolution of the albedo and element composition maps differ by many orders of magnitude. Thus, we created and implemented an adaptive Gaussian blur and compared it against the traditional standard blurring approach. In this setting, each pixel of the image is blurred proportionally, based on the 2D projection of the Moon. In particular, pixels that fall in an X km radius around the target pixel are used for the blurring. This way, the resolution of the albedo map can be adjusted to match the resolution of the element composition maps. The customization of the variance/covariance matrix for the Gaussian filter presented a challenge, given its dependence on specific application scenarios. Accordingly, next, we present the difference between the adaptive and the standard filter (see Figure~\ref{fig:adaptive_blurring}) and additionally a map of how much sigma is applied to each pixel of the image (see Figure~\ref{fig:sigma}).

\begin{figure*}[!ht]
\includegraphics[width=1\textwidth]{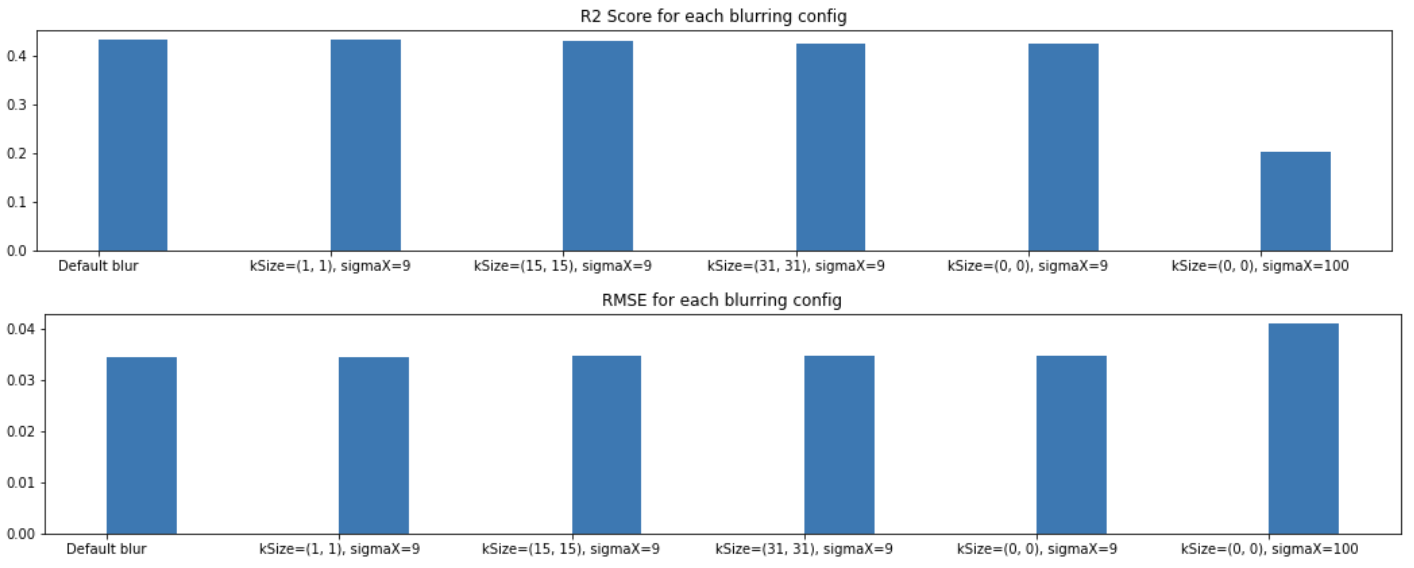}
\centering
\caption{R2 score and root-mean-square error for each blurring configuration}
\label{fig:r2_score}
\end{figure*}

In the interest of objectivity, quantitative metrics were employed to ascertain the efficacy of each blurring method. Specifically, in Figure~\ref{fig:r2_score} we present the R-squared (R2) score and root-mean-square error (RMSE) which were calculated and compared for each experimental setting.

Based on these empirical evaluations, optimal parameter values were identified. For subsequent experiments, we settled on a sigma value of 9 and a kernel size of (0,0), thus establishing a robust methodological foundation for future endeavors.

\begin{figure*}[!ht]
\includegraphics[width=1\textwidth]{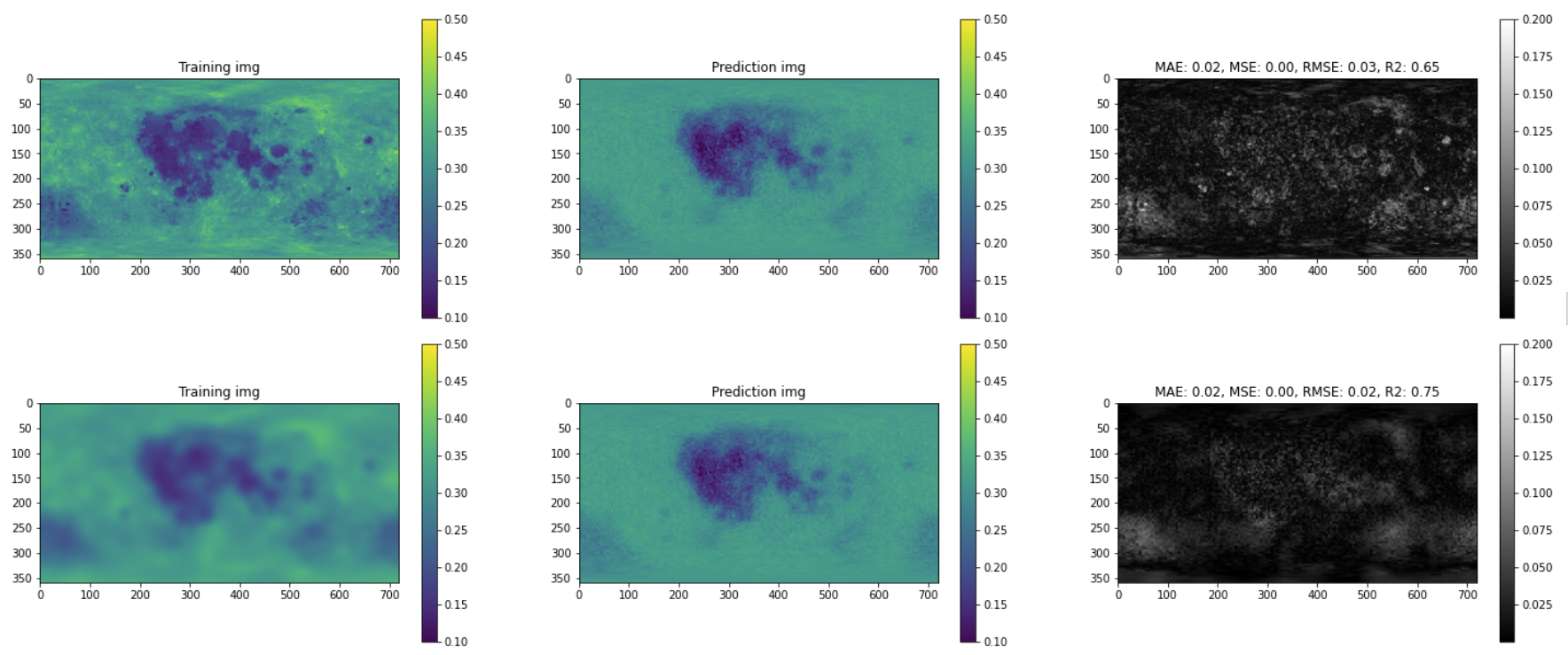}
\centering
\caption{Predicted images in the original and blurred Moon}
\label{fig:predicted_images_original_blurred}
\end{figure*}

Finally, in order to validate the effectiveness of the chosen blurring technique, a regression model was trained on a subset of the surface to generate albedo maps of the Moon in both its original and blurred incarnations. In particular, the inputs of the model are the chemical elements and the albedo is the target. The final best performing model was the extreme gradient boosting regression model (the corresponding learning objective is regression with squared loss, $learning\_rate$ = 0.1, $n\_estimators$ = 30, $max\_depth$ = 5). The results are presented in Figure~\ref{fig:predicted_images_original_blurred}. It serves not only as a testament to the efficacy of the blurring methodology but also as a cornerstone for ongoing and future research in this domain.

\subsection{Interactive Analyzer}

In the concluding phase of our study, we designed an {interactive analytical tool}\footnote{The open source code can be found at
\url{https://github.com/ML4SCI/MLMapper/tree/main/Lunar_Prospector/ML_for_Planetary_Albedo_Sofia_Strukova}} in order to be able to meticulously identify and highlight discrepancies between the predicted and actual image data. This tool augments our methodological framework by providing a dynamic and nuanced visualization medium that enables real-time error assessment.

\begin{figure*}[!ht]
\includegraphics[width=1\textwidth]{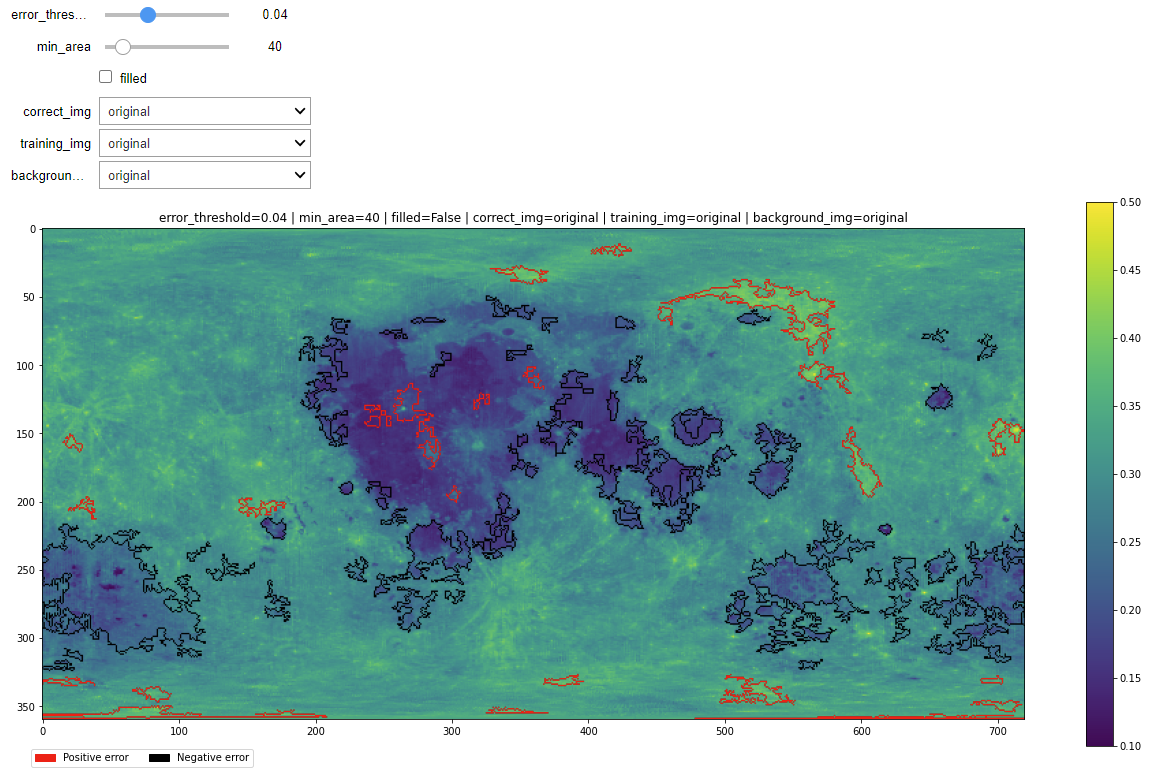}
\centering
\caption{An example of the interactive analyzer (error threshold = 0.04, minimal area = 40, correct image = original, training image = original)}
\label{fig:interractive_analyzer}
\end{figure*}

The interactive analyzer (see Figure~\ref{fig:interractive_analyzer}) displays contours around regions of pixels whose predicted error value is higher than a given threshold. The contours are color-coded for interpretive convenience. Specifically, red contours signify a positive error, while black contours denote a negative error. The tool allows adjustments of the error threshold, a minimum area of the regions, as well as other parameters to configure how the image is displayed. 

\subsubsection{User-Configurable Parameters}

The tool offers an array of adjustable parameters to enhance the granularity of its analyses:

\( \text{mask\_threshold} \): Only consider those pixels whose prediction error is higher than this.

\( \text{min\_area} \): Discard contours whose area is smaller than this.

filled: Fills the contours.

\( \text{correct\_img} \): Image to be used as the correct one for computing the error.

\( \text{training\_img} \): Image to be used to train the ML model.

\( \text{background\_img} \): Image to be displayed in the background.
 
\section{Discussion}
\label{sec:discussion}

\subsection{Observations}

Initial observations based on the results presented above suggest that certain contours can be readily accounted for by intrinsic regional properties. For instance, some areas - such as the Shackleton Crater-display unusually high reflectance relative to the adjacent south polar regions~\cite{Zuber2012}. A considerable proportion of error contours appear to coalesce around young craters and their associated ray systems, hinting at a possible correlation between prediction error and geological age, a topic elaborated upon in the subsequent section of this paper.

Furthermore, we delve into a nuanced error analysis leveraging the capabilities of the interactive tool. Our preliminary observations suggest that prediction errors are not random but correlate with certain geological features and elemental compositions. For instance, errors appear more prominent in regions such as young craters or areas with specific elemental compositions that are not adequately captured in the existing lunar datasets.

This work can be further used as a base for predicting relationships between albedo and chemical composition on other airless bodies in order to obtain precise results and support ongoing and future missions. Mercury is a logical location at which to extend this work. The albedo of the Moon is generally much lower than that of Mercury, but the albedo of the lunar maria is similar to that of Mercury, despite Mercury’s low surface iron centrations~\cite{WEIDER2014170}. This suggests a more significant role for space weathering on Mercury, which can be quantified via application of the ML techniques described here.

\subsection{Optical Maturity}

Optical maturity measures how long material has been on the lunar surface, exposed to the harsh space environment. Space exposure can, among other things, darken a material. Thus, two chemically-equivalent rocks with a different exposure age can have different albedo, meaning that bright material may simply be young. This offers the possibility of using the error map to identify younger optically immature materials independently from previously established techniques, as well as identifying young surface regions at a spatial scale at which existing techniques (e.g. crater counting) may not be reliable.

The Lunar Orbiter Laser Altimeter (LOLA) is an instrument on the payload of NASA's Lunar Reconnaissance Orbiter spacecraft (LRO) [1], which was designed to measure the shape of the Moon by measuring precisely the range from the spacecraft to the lunar surface, and incorporating precision orbit determination of LRO, referencing surface ranges to the Moon's center of mass. LOLA has five beams and operates at 28 Hz, with a nominal accuracy of 10 cm. Its primary objective is to produce a global geodetic grid for the Moon to which all other observations can be precisely referenced.

\begin{figure*}[!ht]
\includegraphics[width=1\textwidth]{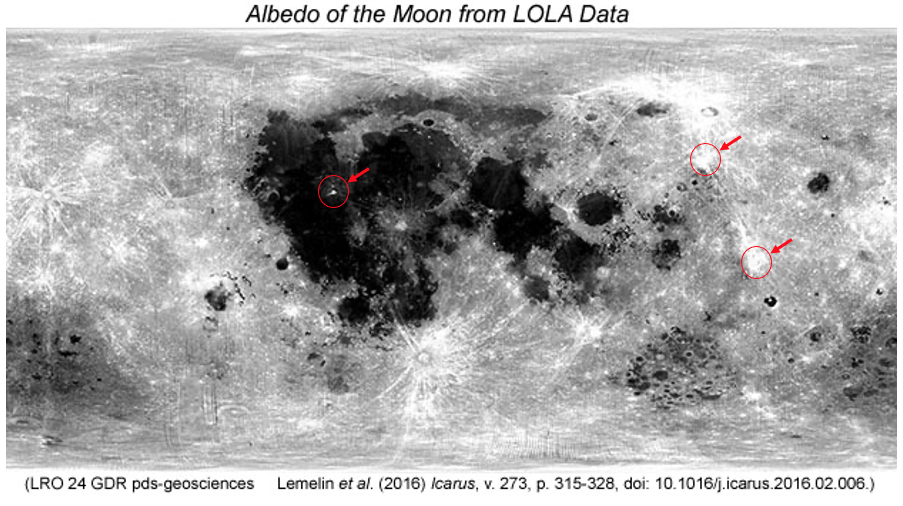}
\centering
\caption{Potential light regions explained by low optical maturity}
\label{fig:potential_light_regions_explained_by_low_optical_maturity}
\end{figure*}

From an optical albedo map, an optical maturity map and an optical maturity map, we can notice that the dark mare is absent, and this fact was expected since here, the track composition is not taken into account. This breakdown between composition and optical maturity allows us to conclude that some of the represented regions indeed have low optical maturity, which we highlight in Figure~\ref{fig:potential_light_regions_explained_by_low_optical_maturity}.

\subsection{Limitations}

We have faced several limitations during this work. Firstly, it is worth noting that specific elements not represented in the Lunar Prospector dataset — namely Iron (Fe), Titanium (Ti), Potassium (K), and Thorium (Th) — might contribute to the observed prediction errors. These elements collectively constitute no more than 20\% of the total chemical inventory at any given location, thereby indicating potential areas for future research.

In addition to elemental representation, another important limitation to consider involves the inherent spatial resolution disparities between the albedo and element maps. While our methodology employs Gaussian blurring techniques to harmonize these differences, the process itself is an approximation and may introduce its own set of artifacts. Furthermore, the choice of the Extreme Gradient Boosting Regression Model, although empirically validated, may still encompass biases or limitations intrinsic to the model itself. For example, the model's configuration and hyperparameters were optimized for a specific set of data and may not be universally applicable for varying lunar terrains or other celestial bodies. As such, both the resolution harmonization and model choice represent avenues warranting additional scrutiny in future research.

\section{Conclusions}
\label{sec:conclusions}

We analyzed the chemical composition and albedo of the Moon using ML methods with the goal of characterizing the relationship between albedo and surface chemistry. Because this relationship is well known on the Moon, this work served to verify the validity of our ML-based approach. By making the albedo maps of comparable resolution to the element maps and removing the spatial resolution difference, we made the data more comparable for training and prediction. This was helpful to minimize potential artifacts that arise from the resolution effects. Our resulting predicted albedo map, and difference from the measured abledo, revealed locations where chemistry is not the dominant effect, for example in optically immature regions.

As a part of an additional experiment, we made a prediction including the surrounding pixels. We achieved slightly better results, but more exploration is needed on the methods that could be used in order to implement this feature and to prove that this setting would benefit the overall results. Moreover, we see potential in exploring the opportunities provided by Toolkit for Multivariate Data Analysis with ROOT and its features applied to high-energy particle physics.

\section{Acknowledgments}

Part of this work was developed during the Google Summer of Code (GSoC) 2021 program as part of the ML4SCI organization.

\bibliographystyle{unsrt}  
\bibliography{references}

\end{document}